\documentstyle[preprint,aps]{revtex}
\tightenlines 

\begin{document}

\title{Coherent Superposition States as Quantum Rulers}
\author{T.C.Ralph}
\address{Centre for Quantum Computer Technology, Department of Physics,
\\ The University of Queensland, \\ St Lucia, QLD 4072 Australia\\
 E-mail: ralph@physics.uq.edu.au}
\maketitle

\begin{center}
\scriptsize (September 2001)
\end{center}

\begin{abstract}

We explore the sensitivity of an interferometer based on a quantum 
circuit for coherent states. We show that its
sensitivity is at the Heisenberg limit. Moreover we show that this 
arrangement can measure very small length intervals.

\end{abstract}

\vspace{10 mm}


There exists a well known isomorphism between interferometers and
basic quantum processing circuits. In particular the
circuit comprising a Hadamard gate followed by a phase gate and then
a second Hadamard gate is equivalent to a single photon, optical interferometer
with a phase shift in one arm (see Fig.1). Historically this
observation has helped to identify candidate quantum circuits.
An alternative viewpoint is to consider the efficacy of such quantum 
circuits in performing more traditionally interferometric tasks 
\cite{lith,clock,weakf}.

We have recently proposed an efficient quantum computation scheme 
based on a coherent state qubit encoding \cite{ral01}. In this paper we
investigate how sensitively distance measurements can be made using
the equivalent of the circuit in Fig.1(a) when realized using this new
scheme. We find that its sensitivity to small perturbations in length
is at the Heisenberg limit. Further more we find that its sensitivity
in measuring small {\it length intervals} is also at the Heisenberg
limit. We refer to this effect as a {\it quantum ruler}.

Our logical qubits are encoded as follows: the zero state is the
vacuum, $|0 \rangle_{L}=|0 \rangle$, and the one state is the
coherent state of amplitude $\alpha$, $|1 \rangle_{L}=|\alpha
\rangle$. We assume that the coherent amplitude is real and that
$\alpha>>1$. Note that this qubit encoding is distinct from other 
quantum circuit \cite{coc99,kim01} and interferometric \cite{rice} 
proposals. We begin by investigating the sensitivity of the
idealized circuit of Fig.1(a) using our coherent state 
qubit encoding and comparing
this with the sensitivity of a standard interferometer with
a squeezed vacuum input 
. We then introduce a physical realization of
the quantum circuit and consider some more practical issues.

Consider the case of the logical zero state, ie the vacuum, entering
the first Hadamard gate. The effect of a Hadamard gate is to
produce the following transformations in the logical basis:
\begin{eqnarray}
|0 \rangle_{L} & \to & {{1}\over{\sqrt{2}}}(|0 \rangle_{L}+
|1 \rangle_{L}) \nonumber\\
|1 \rangle_{L} & \to & {{1}\over{\sqrt{2}}}(|0 \rangle_{L}-|1 \rangle_{L})
\label{H}
\end{eqnarray}
Thus the state of the optical field after the first Hadamard gate is
\begin{eqnarray}
 {{1}\over{\sqrt{2}}}(|0 \rangle +|\alpha \rangle )
\label{H1}
\end{eqnarray}
Now consider small changes in path length (ie phase shifts) around
an integral number of wavelengths ($\lambda$)
between the two Hadamard gates. Propagation over a distance $\Delta$
can be modelled by the unitary operator $\hat U(\theta)=\exp{i \theta \hat
a^{\dagger} \hat a}$ where $\theta=2 \pi \Delta/\lambda$. The
effect of propagation on an arbitrary qubit $|\beta \rangle$, where
$\beta=0$ or $\alpha$, is obtained by examining the overlap
\begin{eqnarray}
\langle \beta | \hat U(\theta) |\beta \rangle & = &
\langle \beta |\beta (\cos{\theta}+i \sin{\theta})
\rangle \nonumber\\
 & = & \exp[-\beta^{2}(1-\cos{\theta}-i \sin{\theta})] \nonumber\\
  & \approx & \exp[i \theta \beta^{2}]
\label{H2}
\end{eqnarray}
where the approximate final result is true in the limit that the
length is small enough that $\theta^{2} \alpha^{2}<<1$ but that
$\alpha$ is sufficiently large that $\alpha^{2} \theta$ is of order $1$.
Eq.\ref{H2} implies that under these conditions $\hat U(\theta)
|\beta \rangle \approx \exp[i \theta \beta^{2}] | \beta \rangle$ and
thus propagation over short distances constitutes a phase gate for this
system:
\begin{equation}
\hat U(\theta) (|0 \rangle +|\alpha \rangle ) \approx
|0 \rangle +e^{i\theta \alpha^{2}}|\alpha \rangle
\label{P}
\end{equation}
Hence the effect of propagation through the entire circuit is given
by
\begin{eqnarray}
|\phi \rangle_{out} & = & \hat H \hat U(\theta) \hat H |0 \rangle
\nonumber\\
& \approx &
 {{1}\over{\sqrt{2}}}((1+e^{i\theta \alpha^{2}})|0 \rangle +
 (1-e^{i\theta \alpha^{2}})|\alpha \rangle)
\label{HPH}
\end{eqnarray}
Clearly the output state is changed as a function of the propagation
distance between the Hadamard gates. We now calculate the sensitivity
to that change.

If no perturbation of the length around an integral number of
wavelengths occurs then the output state will certainly be the
vacuum. Thus the signal strength corresponds to the probability of
finding the output in the state $|\alpha \rangle$. The measurement
noise is the probability that we none-the-less obtain the vacuum
state, $|0 \rangle$ at the output. The signal to noise ratio for
measuring small fluctuations in length is hence
given by
\begin{eqnarray}
S/N = {{| \langle \alpha | \phi \rangle_{out}|^{2}}\over {
| \langle 0 | \phi \rangle_{out}|^{2}}}\nonumber
& \approx & {{V_{\theta} \alpha^{4}}\over{4}} \nonumber\\
& = & V_{\theta} \bar n^{2}
\label{snr}
\end{eqnarray}
Here $V_{\theta}$ is the power in the distance fluctuations and $\bar
n$ is the average photon number in the cat state between the Hadmard
gates ($\bar n=\alpha^{2}/2$).

We now compare the sensitivity of the coherent state quantum circuit to
that of a standard interferometer using a squeezed light input. We
consider the scheme originally proposed by Caves \cite{cav} 
. A beam in a
coherent state with a real amplitude $\beta$ is injected into
one input port 
of an interferometer whilst a phase squeezed vacuum is
injected into the other input port 
. We assume the interferometer is balanced
(equal path lengths in each arm) and consider the null output port.
Small length fluctuations couple into the phase quadrature
of this port. Thus we perform balanced homdyne detection
of the phase quadrature, $X^{-}$, of the null output port. For small length
fluctuations we obtain
\begin{equation}
X^{-} \approx X_{a}^{+} {{\theta}\over{2}} + X_{b}^{-}
\end{equation}
where $X_{b}^{-}$ is the phase (ie the squeezed) quadrature of the
squeezed vacuum and $X_{a}^{+}$ is the amplitude quadrature of the 
coherent input. The signal to noise is then given by
\begin{eqnarray}
S/N & = & {{(\beta^{2}+1) V_{\theta}}\over{4 V_{b}^{-}}}\nonumber\\
& \approx & {{V_{\theta} \bar n^{2}}\over{4}}
\label{snr2}
\end{eqnarray}
where $V_{b}^{-}$ is the noise power in the squeezed quadrature of the
squeezed vacuum. In obtaining the final result in terms of the average
photon number we have assumed that there is equal power in the
coherent beam and the squeezed vacuum and that the squeezed vacuum is
strongly squeezed ($V_{b}^{-}<<1$).

We see that the signal to noise's scale in the same way as a function
of photon number for the two systems. This corresponds to an
amplitude sensitivity which scales as $1/\bar n$, ie the Heisenberg
limit. Thus both systems perform at the ideal limit set by the
uncertainty relations \cite{hei54}. 
The factor of four increase in signal to noise
achieved by the quantum circuit may not be significant. When we
examine a physical realization later in this paper, and keep track of
all resources required, we will find this advantage disappears.

On the other hand there is a significant difference in the way the
increased sensitivity is reached in the two systems which makes the
quantum circuit more versatile. In the squeezed state
interferometer the increase in sensitivity arises from the decrease
in background noise in the measurement. However in the coherent state
circuit the increase is due to a decreasing fringe spacing as
the amplitude of the cat is increased. This means, that as $\alpha$ is
increased, smaller and smaller length intervals can be resolved with a
sensitivity at the Heisenberg limit. This effect is similar to that
recently proposed for increasing lithographic resolution \cite{lith}
and earlier interferometric proposals \cite{burn} based on number 
state superpositions. Increasing the
power in the cat state is effectively the same as increasing the
frequency of the light in a standard interferometer, and thus
decreasing the fringe spacing. We believe this effect could have
important applications.

We now consider a physical implementation of our quantum circuit.
This is shown schematically in Fig.2. A coherent state phase reference 
beam is divided at a 50:50 beamsplitter. One of the beams is sent to  
a ``cat-state maker'' of some
kind , which produces the state given by Eq.\ref{H1}, in phase with 
the reference beam.
Such a device is not trivial of course, though some limited
success has been achieved in making such devices experimentally \cite{exp}. 
Also the cat-state 
maker need not necessarily be deterministic. In principle one could
imagine building up a resource of the required cat-states which are
then fed into the interferometer when the measurement is required. A
number of non-deterministic schemes for producing cat states have been
proposed \cite{song}.

The cat-state maker performs the role of the first Hadamard gate in 
the idealized circuit (Fig.1(a)). The cat-state beam is then passed along the 
path whose distance is to
be measured. In order to implement the second Hadamard gate we use
the scheme proposed in Ref.\cite{ral01}. A second cat state,
identical to the first, and phase locked to the second coherent reference beam, 
is weakly mixed with the beam at a highly
reflective beamsplitter. A surprizing result from Ref. \cite{ral01} 
is that such a beamsplitter, with reflectivity $\cos^{2}\phi$ where $\phi^{2}
\alpha^{2}<<1$ but $\phi \alpha^{2}=\pi/2$, will act as a
control sign gate \cite{note1} for our coherent state qubits. 
As a result if output state $c$ in Fig.2 is
measured in the ``cat-basis'' (see below) 
and is found to be in the same cat-state
as was injected, then the required Hadamard transformation is
implemented onto beam $d$.
Alternatively if the output is found in the (near)
orthogonal state $1/(\sqrt{2})(|0 \rangle -|\alpha \rangle )$, then
the output state is a bit flipped version of the Hadamard gate. Homodyne 
amplitude measurements are performed on output $d$ and the data is 
collected in two bins according to the results of the cat-basis 
measurements.

The cat basis measurements can be made by combining
displacements and photon number measurements \cite{ral01}. The 
procedure is: first displace by
$-\alpha/2$. This transforms our ``0'', ``$\alpha$'' superposition into
``$\alpha/2$'', ``$-\alpha/2$'' superposition:
\begin{equation}
D(-\alpha/2)1/\sqrt{2}(|0\rangle \pm|\alpha
\rangle)=1/\sqrt{2}(|-\alpha/2\rangle \pm|\alpha/2 \rangle)
\end{equation}
These new states are parity eigenstates. Thus if photon number is
measured then an even result indicates detection of the state
$1/\sqrt{2}(|\alpha/2\rangle +|-\alpha/2 \rangle)$ and therefore
$1/\sqrt{2}(|0\rangle +|\alpha \rangle)$ whilst similarly an odd
result indicates detection of $1/\sqrt{2}(|0\rangle -|\alpha
\rangle)$ as can be confirmed by direct calculation.

Notice our physical
implementation requires two cat states as resources. Clearly this
other resource should be included in calculating the signal to noise
in terms of the photon number. The extra factor of 2 will then make
the results for the squeezed state and cat schemes equivalent in this 
realization.

Having a physical implementation we can now make realistic
calculations to confirm the efficacy of the measurement
protocol for finite values of $\alpha$. To do this we use the exact
solution for the output field for which no assumptions about the 
magnitude of $\alpha$ have been made. Using the beamspitter 
relationship $|\gamma \rangle_{a} |\beta \rangle_{b} \to |\cos \theta \gamma+i
\sin \theta \beta \rangle_{a} |\cos \theta \beta+
i \sin \theta \gamma \rangle_{b}$, a straightforward calculation 
gives:
\begin{eqnarray}
|out \rangle_{\pm} & = & {{1}\over{2+2 e^{-\alpha^{2}/2}}}
(A_{\pm}|0 \rangle + 
B_{\pm} |i \alpha \sin \phi e^{i \theta} \rangle \nonumber\\
 & & + C_{\pm} |\alpha \cos \phi \rangle + D_{\pm} | \alpha(\cos \phi+
i \sin \phi e^{i \theta} \rangle)
\label{exp1}
\end{eqnarray}
where 
\begin{eqnarray}
A_{\pm} & = & {{1}\over{\sqrt{2 \pm 2 e^{-\alpha^{2}/2}}}} (\langle 
0| \pm \langle \alpha|) |0 \rangle \nonumber\\
B_{\pm} & = & {{1}\over{\sqrt{2 \pm 2 e^{-\alpha^{2}/2}}}} (\langle 
0| \pm \langle \alpha|) |\alpha \cos \phi e^{i\theta} \rangle \nonumber\\
C_{\pm} & = & {{1}\over{\sqrt{2 \pm 2 e^{-\alpha^{2}/2}}}} (\langle 
0| \pm \langle \alpha|) |i \alpha \sin \phi \rangle \nonumber\\
D_{\pm} & = & {{1}\over{\sqrt{2 \pm 2 e^{-\alpha^{2}/2}}}} (\langle 
0| \pm \langle \alpha|) |\alpha(\cos \phi+i \sin \phi e^{i \theta}) \rangle
\label{exp2}
\end{eqnarray}
and $\phi=\pi/(2 \alpha^{2})$. The state overlaps can be calculated using
the relationship \cite{Wal94} $\langle \tau|\alpha \rangle =
\exp[-1/2(|\tau|^{2}+|\alpha|^{2})+\tau^{*} \alpha]$. 
The $+$ subscript refers to the situation where a plus cat 
(ie: $|0 \rangle +|\alpha \rangle$) is found at output 
$c$ and the $-$ to the case where a minus cat is found 
(ie: $|0 \rangle -|\alpha \rangle$). We then calculate 
\begin{equation}
P_{\pm}={{1}\over{\sqrt{\pi}}}\int_{-\infty}^{\alpha/2} 
|\langle x'|out \rangle_{\pm}|^{2} d x'
\label{wv}
\end{equation}
where $\psi_{out}=\langle x'|out \rangle_{\pm}$ is the amplitude 
quadrature wave function of the output field and can be calculated 
using 
\begin{equation}
    |\langle x'|\gamma 
    \rangle|^{2}=e^{-({{\gamma+\gamma^{*}}\over{2}}-x')^{2}}
\end{equation}
Eq.\ref{wv} gives the probability that a measurement of the amplitude 
quadrature of output beam $d$ gives a result lying below 
$\alpha/2$. This we consider a ``0'' result. 
When a plus cat is found at output $c$ we 
label this result $P_{+}$. When a minus cat is found at output $c$ we 
label the result $P_{-}$. The two probabilities show fringes as a 
function of $\theta$ but they are $\pi/2$ out of phase. Note that this means 
that without the cat-basis measurements to distinguish the two cases 
the fringes would be completely washed out. 

With the cat basis binning of the results we are able
to form the following function: $(P_{-}-P_{+}+1)/2$, which corrects 
for the bit flip between the results. This is 
plotted for various values of $\alpha$ in Fig.3. The width of the 
middle fringe scales as $1/\alpha^{2}$ between the three graphs (note 
changing axis scale). This indicates sensitivity at the Heisenberg 
limit. 

The quantum ruler effect is also clear. As $\alpha$ increases, a 
number of high visibility, narrowly spaced fringes emerge, which could 
enable very short length intervals to be accurately measured. As an 
example suppose our laser wavelength is 1$\mu m$. In a standard 
interferometer this would enable length intervals of $0.5 \mu m$ to 
be stepped off. The use of squeezing would increase the precision of 
our measurements but would not change the length scale. However 
using the cat-state interferometer with an $\alpha$ of 20 (Fig.3(c)) 
leads to the fringe separation being reduced to $1.25 nm$.

We have introduced an interferometer based on a recently introduced 
quantum circuit for coherent states and their superposition. We have 
shown that this arrangement has a sensitivity at the Heisenberg limit 
and also displays a quantum ruler effect which could be used to 
resolve precisely very small length intervals. As well as possible 
applications in metrology the experiments suggested here may serve as 
an initial testing ground for coherent state quantum circuits.

We acknowledge useful discussions with G.J.Milburn and W.J.Munro. 
This work was supported by the Australian Research Council.

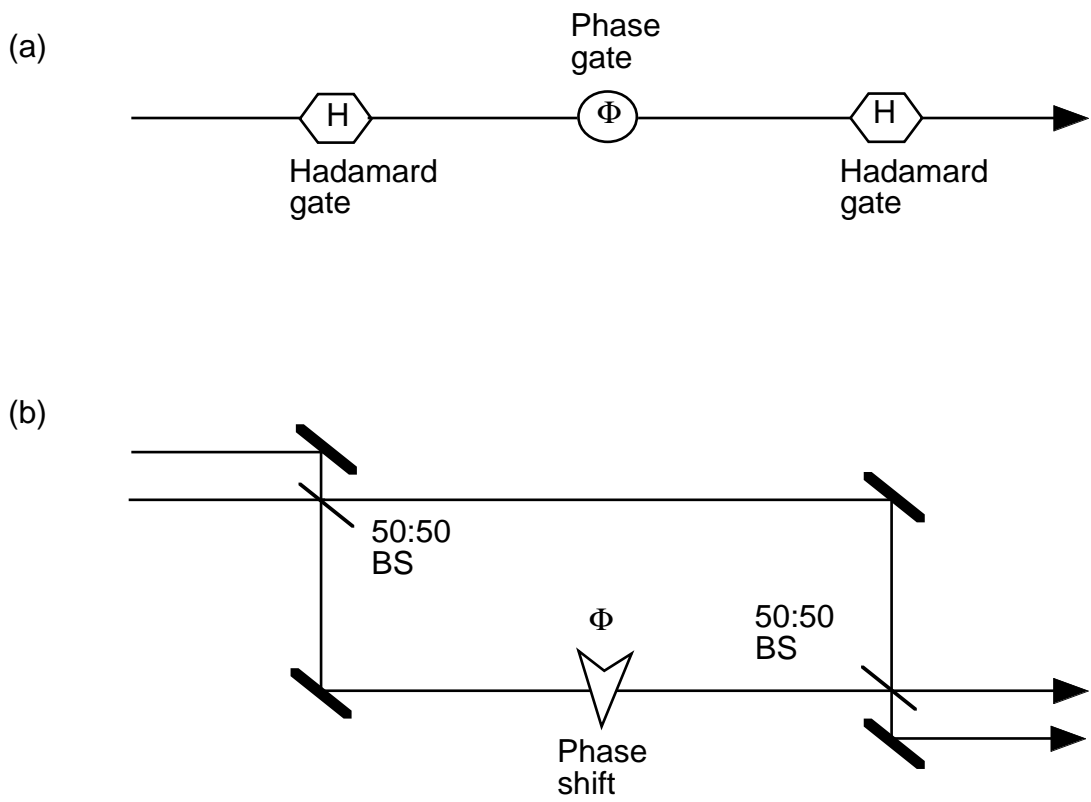
\begin{figure}
 \caption{Schematics of quantum circuit (a) and optical 
 interferometer (b). If a single photon is incident on the 
 interferometer then the description of the path of the photon is 
 mathematically equivalent to the description of the state of the 
 qubit in the quantum circuit with the beamsplitters (BS) playing the 
 role of the Hadamards and the phase shift that of the phase gate.}
\end{figure}


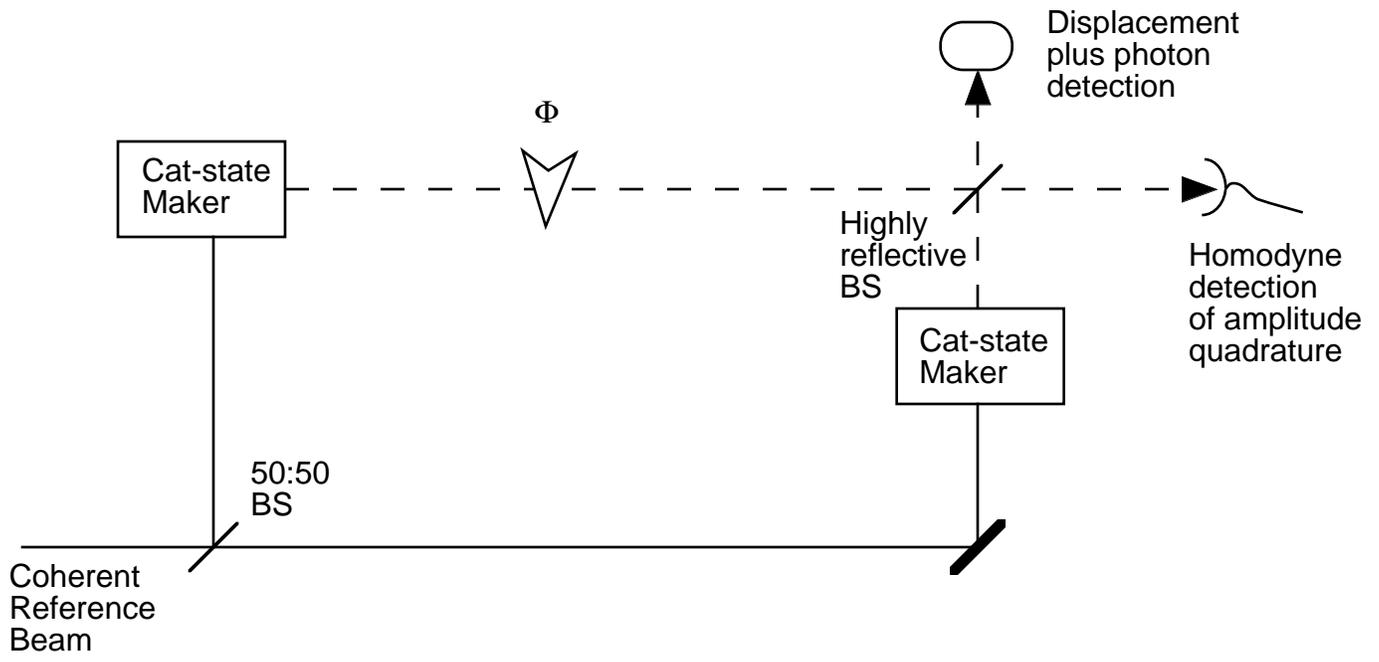
\begin{figure}
 \caption{Schematic of a physical realization of the quantum circuit of 
 Fig.1(a) using coherent state encoding. Solid lines are used to 
 indicate coherent beams whilst dashed lines are beams that in 
 general are in
 superposition states.}
\end{figure}

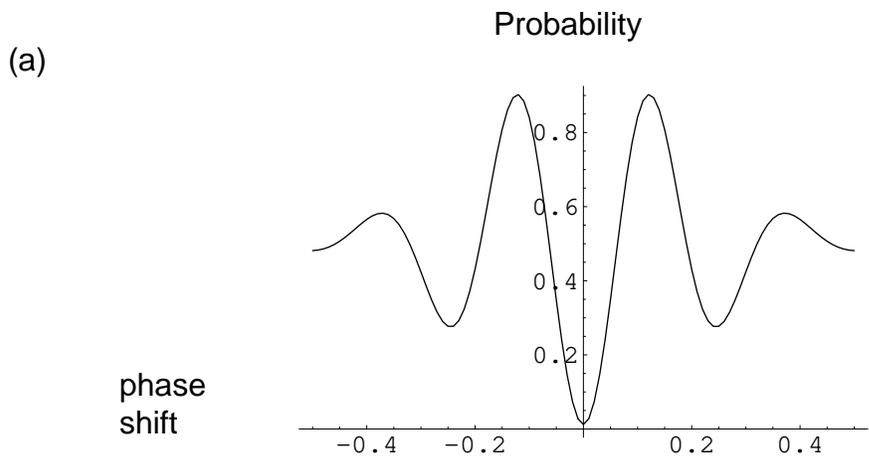
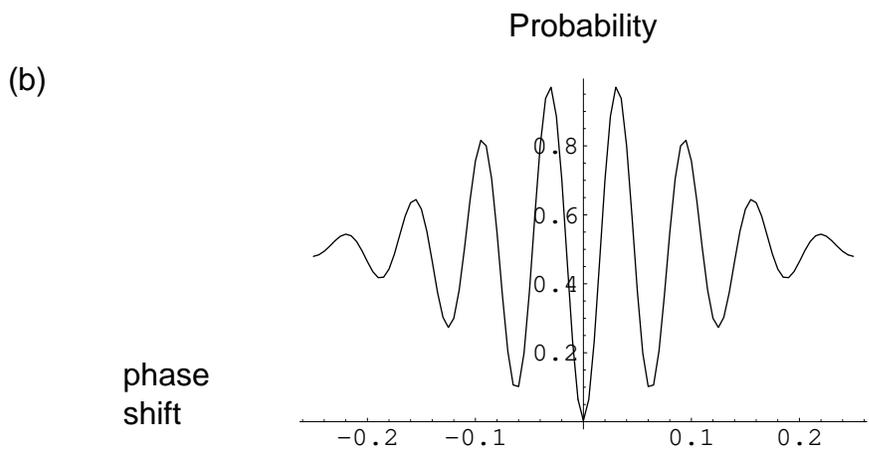
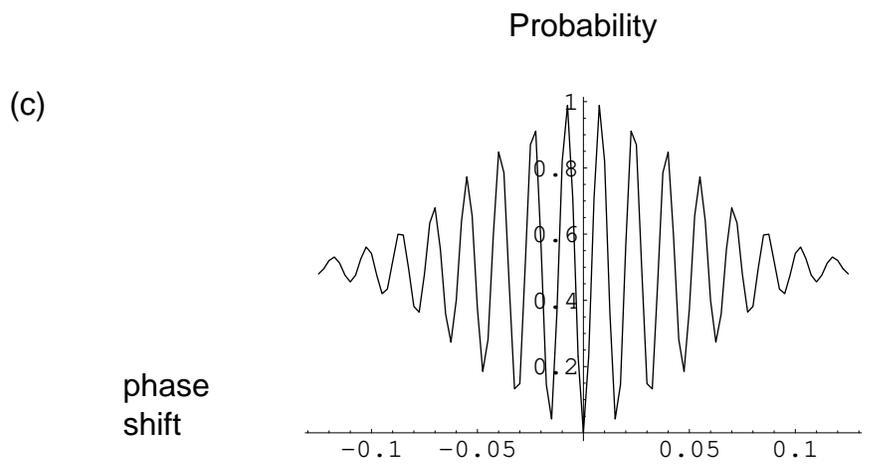
\begin{figure}
 \caption{Probability of obtaining ``0'' result as a function of the 
 phase shift/distance shift in the interferometer. The coherent 
 amplitude is varied between the three graphs. In (a) $\alpha=5$, in 
 (b) $\alpha=10$ and in (c) $\alpha=20$. Note that the scale on the 
 horizontal axis of each graph is scaled by $1/\alpha$.}
\end{figure}


\begin{thebibliography}{99}

\bibitem{lith} A.~N.~Boto, P.~Kok, D.~S.~Abrams,
 S.~L.~Braunstein, C.~P.~Williams and J.~P.~Dowling, 
\prl {\bf 85}, 2733 (2000).

\bibitem{clock} V.~Giovannetti, S.~Lloyd, L.~Maccone, and F.~N.~C.~Wong,
\prl {\bf 87}, 117902 (2001).

\bibitem{weakf} W.~J.~Munro, K.~Nemoto, G.~J.~Milburn, 
S.~L.~Braunstein, quant-ph/0109049 (2001).

\bibitem{ral01} T.~C.~Ralph, W.~J.~Munro and G.~J.~Milburn, submitted 
to Nature (2001) preprint available on request.

\bibitem{coc99} P.~T.~Cochrane, G.~J.~Milburn and W.~J.~Munro, \pra 
{\bf 59}, 2631 (1999)

\bibitem{kim01} H.~Jeong and M.~S.~Kim, quant-ph/0109077 (2001).

\bibitem{rice} D.~A.~Rice, G.~Jaeger and B.~C.~Sanders, \pra {\bf 
62}, 012101 (2000).

\bibitem{cav} C.~M.~Caves, \prd {\bf 23}, 1693 (1981).

\bibitem{hei54} W.~Heitler, {\it The Quantum Theory of Radiation} 
(Oxford University Press, Oxford, 1954).

\bibitem{burn} M.~J.~Holland and K.~Burnett, \prl {\bf 71}, 1355 (1993).

\bibitem{exp} C.~Monroe, D.~M.~Meekhof, B.~E.~King 
and D.~J.~Wineland, Science {\bf 272}, 1131 (1996), M.~Brune, E.~Hagley, 
J.~Dreyer, X.~Maitre, A.~Maali, C.~Wunderlich, J.~M.~Raimond 
and S.~Haroche, \prl {\bf 77}, 4887 (1996), Q.~A.~Turchette, C.~J.~Hood, 
W.~Lange, H.~Mabuchi and H.~J.Kimble, \prl {\bf 75}, 4710 (1995).

\bibitem{song} S.~Song, C.~M.~Caves and B.~Yurke,
\pra {\bf 41}, 5261 (1990), M.~Dakna, T.~Anhut, T.~Opatrny, L.~Knšll and 
D.~-G.~Welsch, \pra {\bf 55}, 3184 (1997).

\bibitem{note1} A control sign gate produces the following logic 
table: $|0 \rangle_{L} |0 \rangle_{L} \to 
 |0 \rangle_{L} |0 
\rangle_{L}$, $|0 \rangle_{L} |1 \rangle_{L} \to |0 
\rangle_{L} |1 
\rangle_{L}$, $|1 \rangle_{L} |0 \rangle_{L} \to |1 \rangle_{L} |0 
\rangle_{L}$ but $|1 \rangle_{L} |1 \rangle_{L} \to -|1 \rangle_{L} 
|1 \rangle_{L}$.

\bibitem{Wal94} D.~F.~Walls and G.~J.~Milburn, {\it Quantum Optics}
(Springer-Verlag, Berlin, 1994).

\end{thebibliography}
\end{document}